\documentclass[10pt]{article}
\usepackage{epsfig}
\usepackage{subfigure,subeqn}
\begin{document}

\title{Slow-light Faraday effect: an atomic probe with gigahertz bandwidth.}

\author{Paul~Siddons*,  Nia~C~Bell, Yifei~Cai,\\ Charles~S~Adams and Ifan~G~Hughes\\ \\ Department of Physics,
Durham University, \\South Road, Durham, DH1~3LE, UK\\\\
paul.siddons@durham.ac.uk}

\date{\today}

\maketitle

\begin{abstract}
\noindent The ability to control the speed and polarisation of light pulses will allow for faster data flow in optical networks of the future.  Optical delay and switching have been achieved using slow-light techniques in various media, including atomic vapour.  Most of these vapour schemes utilise resonant narrowband techniques for optical switching, but suffer the drawback of having a limited frequency range or high loss.  In contrast, the Faraday effect in a Doppler-broadened slow-light medium allows polarisation switching over tens of GHz with high transmission. This large frequency range opens up the possibility of switching  telecommunication bandwidth pulses and probing of dynamics on a nanosecond timescale.  Here we demonstrate the slow-light Faraday effect for light detuned far from resonance.  We show that the polarisation dependent group index can split a linearly polarised nanosecond pulse into left and right circularly polarised components.  The group index also enhances the spectral sensitivity of the polarisation rotation, and large rotations of up to 15$\pi$~rad are observed for continuous-wave light.  Finally, we demonstrate dynamic broadband pulse switching, by rotating a linearly polarised nanosecond pulse from vertical to horizontal with no distortion and transmission close to unity.  

\end{abstract}

The phenomenon of reduced optical group velocity (slow light) is a topic of burgeoning interest~\cite{NatPhoton}. In a slow-light medium, the group refractive index, $n_g$, (the ratio of the speed of light \textit{in vacuo} to the pulse velocity) is many orders of magnitude larger than the phase index, $n$.  Hence  an optical pulse propagates much more slowly than a monochromatic light beam.  Large group indices of $\sim10^7$ are achievable in resonant optical processes, such as electromagnetically induced transparency (EIT), accompanied by a refractive index that is of the order of unity~\cite{Hau1999}. Such large group indices are the result of a rapid change in refractive index over a narrow spectral range, requiring narrowband pulses.  As the relatively small bandwidth limits the ability to produce large fractional delay (the ratio of the delay to the width of the pulse), recent slow-light experiments based on EIT have focussed on light storage~\cite{Lukin00,Scully01,Polzik08}.  Larger bandwidths can be obtained in Doppler-broadened media using self-induced transparency (SIT). SIT has been used to rotate pulses as short as 5~ns~\cite{Gibbs74}.  However, this requires intensities above saturation and the near-resonant character of the process leads to higher order terms in the dispersion resulting in considerable pulse distortion.  Slow light has also been studied in off-resonant optical systems, utilising Doppler-broadened absorption resonances~\cite{Vanner08,Camacho07, Tanaka03}. In such systems, group indices of $\sim10^3$ are achievable, with GHz pulse bandwidths.  This broad bandwidth allows large fractional pulse delays of up to 80~\cite{Camacho07}.  Delaying pulses by more than their widths has applications in data synchronisation and qubit operations for quantum computing~\cite{Gauthier06}.  Another interesting application  of slow light is interferometry, using either monochromatic light sources~\cite{Boyd07cw,Purves06}, where the large dispersion associated with a slow-light medium results in greater phase sensitivity; or polychromatic light, where the large pulse delays can increase the resolution of a Fourier transform interferometer by orders of magnitude~\cite{Boyd07FT}.

The Faraday effect is the rotation of linearly polarised light in a medium with an applied magnetic field parallel to the direction of light propagation.  The rotation is caused by induced circular birefringence: the medium responds differently to left and right circularly polarised light.  The birefringence produces a phase difference between circular field components.  The relative phase shift, $\Delta\phi$, with spectral dependence, $\rm{d}(\Delta\phi)/\rm{d} \omega$ (see Theoretical analysis in Methods), is given by \begin{eqnarray}
\Delta\phi&=&\frac{\omega L}{c} (n^+-n^-)~,\label{eq:1}\\
\frac{\rm{d}(\Delta\phi)}{\rm{d} \omega} & = &  \frac{L}{c}(n^+_g-n^-_g),
\label{eq:phiCirc}
\end{eqnarray}
where $n^+-n^-$ ($n^+_g-n^-_g$) is the differential phase (group) index, i.e., the difference between the indices of $\sigma^+$ and $\sigma^-$ transitions.  The effect of a relative phase shift is to rotate the polarisation angle of the linear input by an amount $\theta=\Delta\phi/2$. Equation~(\ref{eq:phiCirc}) shows how the spectral dependence of the Faraday effect is enhanced by the large group index in a slow-light medium.  Resonant magneto-optical effects have been studied extensively over the past century~\cite{Budker02}.  While the off-resonant Faraday effect has been used to probe semiconductor spin ensembles~\cite{SemCon99} and quantum dots~\cite{QD07}, in atomic systems it has been largely neglected.  In a slow-light medium, large dispersion is encountered with relatively low absorption when the light is detuned from resonance by several Doppler-broadened linewidths (i.e. GHz).  A magnetic field applied to such a medium can produce rotations of greater than $\pi/2$~rad without significant absorption.  As the rotation is controlled by the applied field the effect can be used as a tunable frequency filter~\cite{Menders91}; and for pulsed light it can lead to pulse splitting~\cite{Splitting73}.

In this article we demonstrate the slow-light Faraday effect in Rb vapour by propagating separate beams of continuous-wave (cw) and pulsed laser light, showing that for a modest applied magnetic field and a cell temperature of 100-200$^\circ$C the effect on light propagation can be dramatic.  We show that in addition to pulse splitting the spectral response of the Faraday rotation has increased sensitivity due to the large group index associated with a slow-light medium.  By using a cw beam to measure polarisation rotation as function of frequency, we obtain both the phase and group indices.  Finally, we highlight the large bandwidth of the slow-light Faraday effect and demonstrate controlled rotation of a nanosecond pulse.     

\section*{Propagation of optical pulses}

\begin{figure}[h!]
\centering
\includegraphics[width=8.5cm]{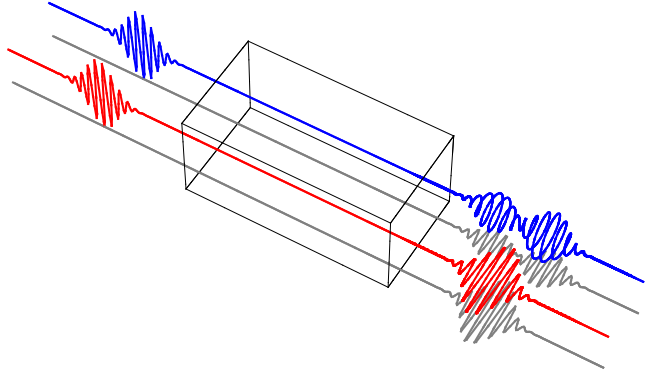}
\caption{An illustration of the Faraday effect for pulsed light.  A representation of two pulsed electric fields, propagating from left to right along the z axis.  The other axes show the electric field magnitudes $E_x$ and $E_y$.  The box marks the position of the vapour cell where the magneto-optical effect occurs.  Both input beams are linearly polarised along the vertical axis. The two cases correspond to different values of either the detuning, magnetic field or atomic density.  The red beam remains linear as its polarisation angle is rotated on exiting the medium, whilst the blue beam is split into two pulses of left and right circular polarisation.}
\label{fig:Efield}
\end{figure}

The properties of pulse polarisation rotation and splitting are illustrated in figure~\ref{fig:Efield}.  For a small relative group index the pulse retains its shape, with its linear polarisation rotated by an amount dependent on the relative phase index. If the relative group index is larger, the pulse splits into two spatially (and temporally) separated circular components of opposite chirality.  By varying the relative refractive indices it is possible to switch dynamically between split or rotated pulses (and also control the magnitude of this rotation).  We show that this is possible by changing a single experimental parameter.  

\begin{figure}[tbh]
\centering
\includegraphics[width=8.5cm]{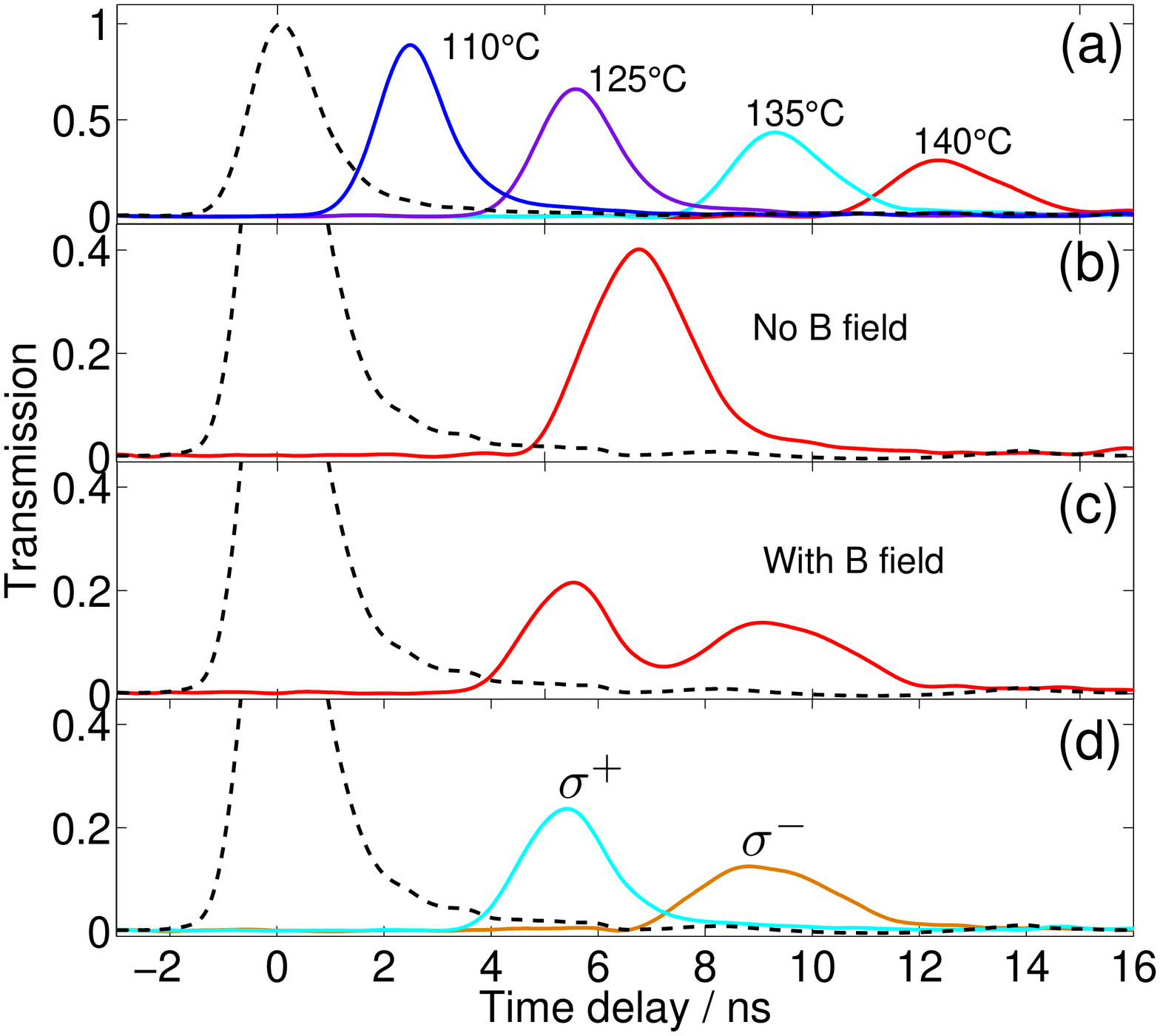}
\caption{Optical pulse propagation in a slow-light medium.  Propagation of an input 1.5~ns pulse through a Rb vapour cell.  Transmission and delay are relative to a non-interacting reference pulse (dashed). (a) Pulse form at various temperatures for a pulse centred at the  frequency of high transmission near the D$_1$ transition centre~\cite{Siddons08}. (b)-(d) Pulse red-detuned from the $^{87}$Rb $F=2\rightarrow F^\prime$ transition by $\sim 3.5$~GHz at a cell temperature of approximately 165$^\circ$C.  A magnetic field of $\sim360$~G is then applied to the linearly polarised pulse in plot (c); (d) shows the left/right circular components of the split pulse.}
\label{fig:pulse}
\end{figure}

The experimental setup for demonstrating the slow-light Faraday effect is described in the Methods section.  The results of pulse propagation through a Rb vapour cell are shown in figure~\ref{fig:pulse}.  The slow-light effect is illustrated in figure~\ref{fig:pulse}(a),  were large fractional delay for a pulse centred between two absorption lines is seen, as observed in previous work \cite{Vanner08,Camacho07,Tanaka03}.  A far off-resonant pulse is shown in \ref{fig:pulse}(b)-(d). In figure~\ref{fig:pulse}(c), we show the effect of applying a magnetic field along the propagation direction.  In this case the pulse is split into a double pulse.  By using a $\lambda/4$ waveplate  it is possible to observe the magneto-optic effect on the $\sigma^\pm$ transitions individually, shown in \ref{fig:pulse}(d).  Comparison of these last two plots shows that the linearly polarised pulse is split into its two circular component as illustrated schematically in figure~\ref{fig:Efield}.  This effect was first observed by Grischkowsky (ref.~\cite{Splitting73}) in a 1~m long cell at 140$^\circ$C with a 8500~G pulsed magnetic field, who recorded a time separation of 25~ns, with a $\sigma^-$ velocity of $c/9$ (compared to our observations of 3.5~ns and $c/37$, respectively).     

\begin{figure}[tbh]
\centering
\includegraphics[width=8.5cm]{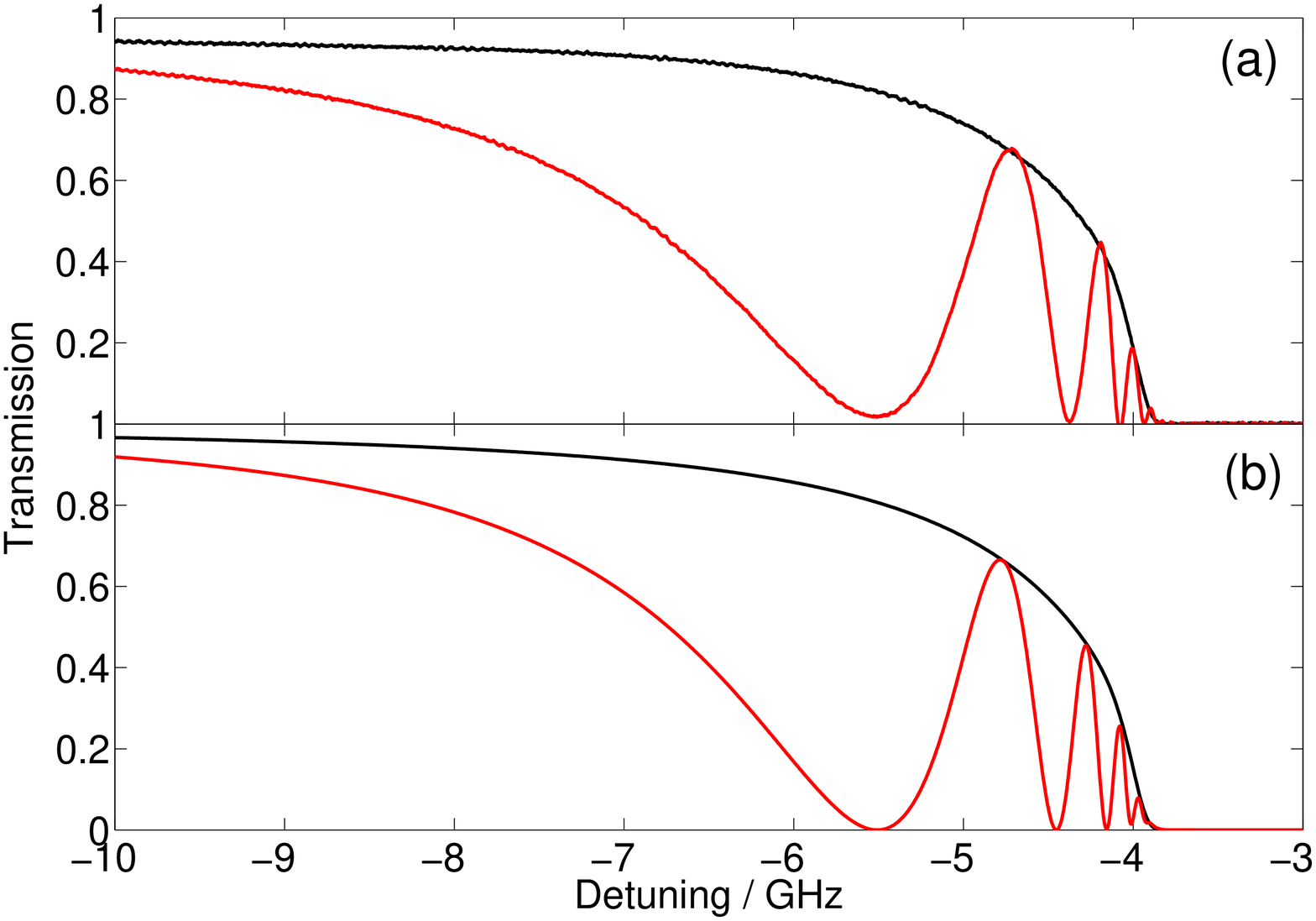}
\caption{Continuous-wave propagation in a slow-light medium.  Transmission of the horizontal component of the cw beam through the vapour cell and polarisation beam splitter at 136$^\circ$C.  The black curve is for zero applied magnetic field, the red is for a field of $\sim50$~G.  
(a) Experimentally measured data and (b) theoretical model.  Zero detuning corresponds to the weighted centre of the D$_1$ line.}
\label{fig:Ix}
\end{figure}

\section*{Spectral dependence of polarisation rotation}

The effect of an applied magnetic field on cw spectroscopy is as dramatic as the effect on pulse propagation.  In figure~\ref{fig:Ix}(a) the conventional Doppler-broadened transmission signal is seen to oscillate between zero and the zero-field transmission when a modest field is applied.  This is a consequence of polarisation rotation due to the Faraday effect.  Large rotations of above $\pi$~rad are seen close to resonance, and even at detunings of many GHz the difference in transmission is noticeable.  The rapid oscillation in the polarisation as a function of frequency has a number of applications such as a polarisation switch, optical isolator or narrowband filter~\cite{Menders91}.  By using isotopically pure Rb cells one could use one isotope to realise a device that works on resonance for the other isotope. Such applications will be the focus of future work.   The experimental data are compared to the predictions of a theoretical model presented in reference~\cite{Siddons08} in figure~\ref{fig:Ix}(b).  The theoretical model assumes a linear Zeeman response to the applied field. In this regime the model predicts both the Doppler-broadened lineshape, and the number and positions of the magneto-optically induced fringes.

\begin{figure}[tbh]
\centering
\includegraphics[width=9cm]{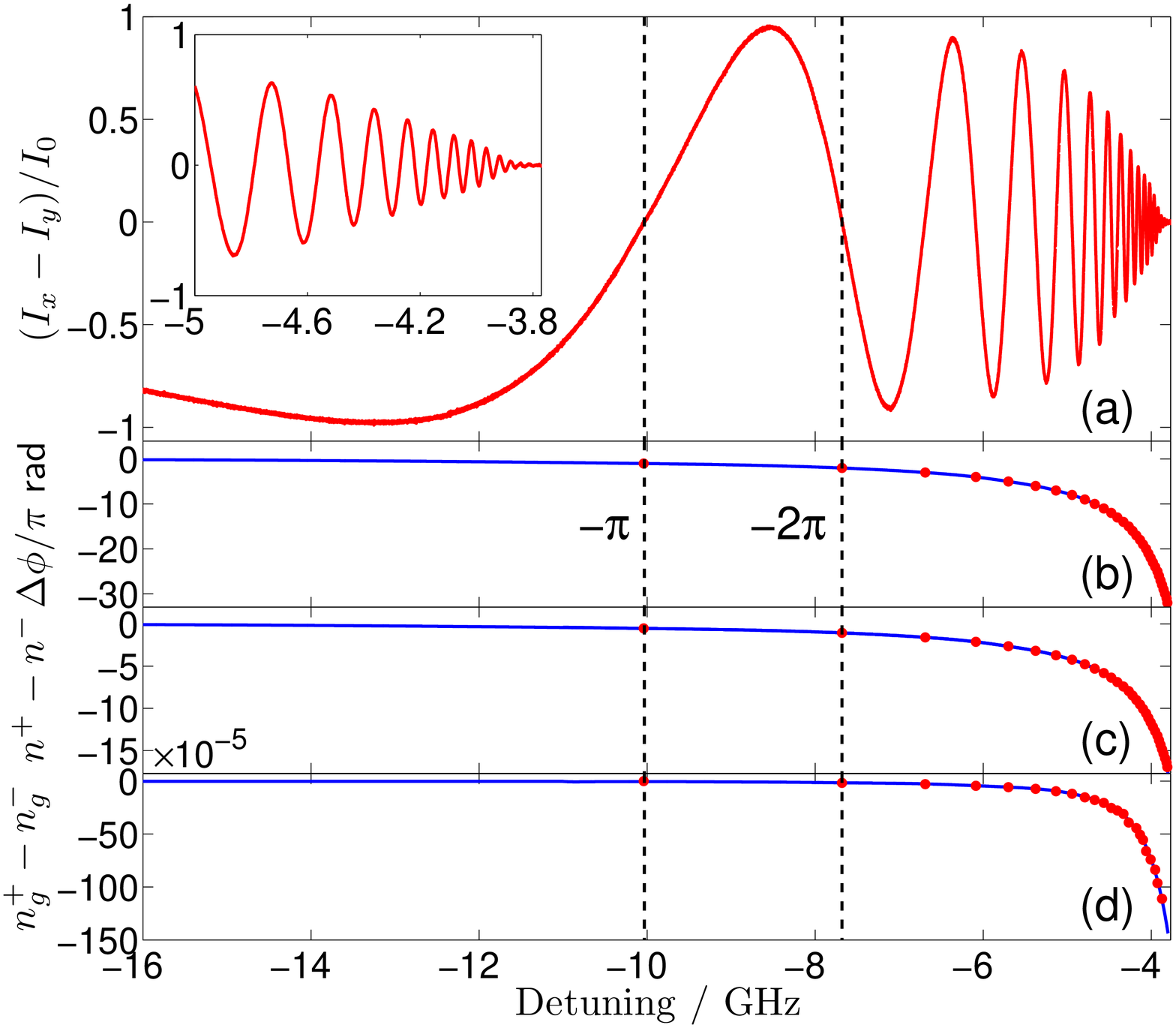}
\caption{Determination of relative phase and refractive indices using a cw beam.(a) The differencing signal obtained by subtracting the two orthogonal linear polarisations for an input polarisation orientated at $\pi/4$~rad to the horizontal, for a vapour temperature of 135$^\circ$C and applied magnetic field of 230~G.  The signal is normalised to the total intensity of the input.  The inset shows a smaller range of detuning.  (b) the relative phase determined from the zero crossing of the differencing signal (red points), with interpolated trendline (blue curve);  (c) Relative refractive index calculated from the measured phase; and (d) the relative group index calculated from the gradient of the phase.  Zero detuning corresponds to the weighted centre of the D$_1$ line.}
\label{fig:Diff}
\end{figure}

\section*{Refractive index measurements}

As the magnetic field of the cell is increased further, the number of fringes increases rapidly. Figure~\ref{fig:Diff} shows experimental data collected using a balanced polarimeter (see Methods), where the differencing signal of vertical and horizontal polarisations is recorded. In this case over 30 fringes are observed before one reaches a detuning sufficiently small that the medium is optically thick.  These large rotations are useful for applications that require multiple $\pi$ rotations.

The large number of fringes observed in figure~\ref{fig:Diff} arises from the same phenomenon that enhances the spectral response in a slow-light interferometer \cite{Boyd07cw} (see Methods). A powerful property of the polarimetry signal is that it allows a direct read-out of the differential refractive and group indices. The signal tends to zero at large negative detuning (not shown) and each subsequent zero corresponds to a $\pi$ phase shift between the left and right circular components. This phase difference, $\Delta\phi$, is plotted in figure~\ref{fig:Diff}(b).  Note that the first zero crossing (corresponding to a $\pi/2$ rotation) occurs at a detuning of $\sim-10$~GHz where the absorption is less than $1\%$. At temperatures of order 200$^\circ$C, the first zero shifts out to beyond $-50$~GHz i.e. $\sim100$ Doppler linewidths. The phase obtained from the zero crossing is used to determine the differential index, Eq.~(\ref{eq:1}) which is plotted in figure~\ref{fig:Diff}(c).  Similarly, the differential group index is determined by the frequency spacing of the zero crossings, Eq.~(\ref{eq:phiCirc}), and is shown in figure~\ref{fig:Diff}(d).  

\section*{Broadband pulse rotation}

\begin{figure}[!h]
\centering
\includegraphics[width=9cm]{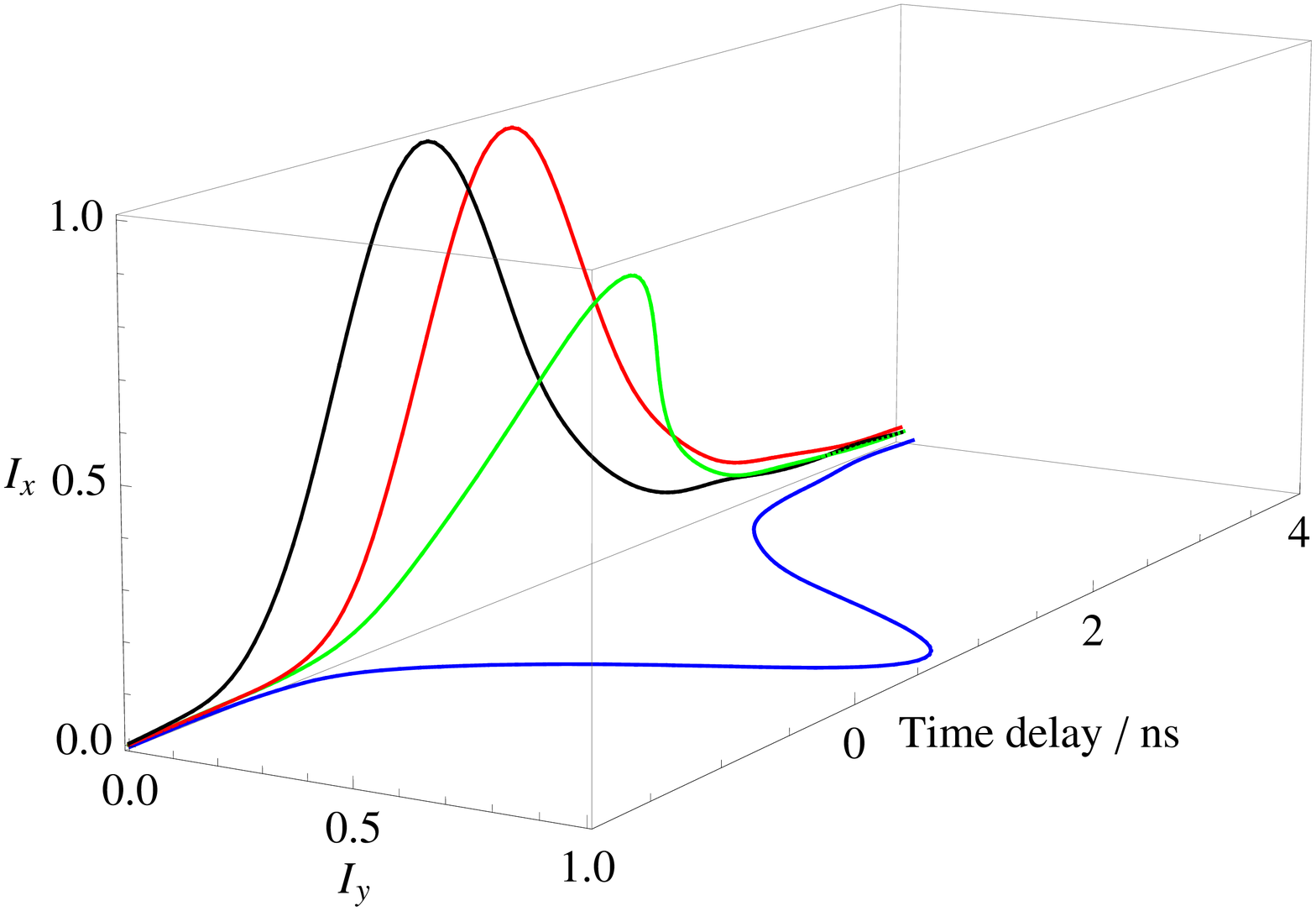}
\caption{Broadband Faraday rotation in a slow-light medium.  An input 1.5~ns pulse initial linearly polarised in the x direction (red) is delayed by 0.6~ns with respect to a non-interacting reference pulse (black), in the absence of an applied magnetic field.  The pulse is red-detuned from the weighted D$_1$ transition centre by $\sim$10~GHz, for a vapour temperature of 135$^\circ$C.  A field of 80~G (green) and 230~G (blue) is applied, showing the pulse rotating into the y direction whilst retaining its linear polarisation and intensity.}
\label{fig:pulserot}
\end{figure}

It can be seen in figure~\ref{fig:Diff}(b) that far from resonance the relative phase evolves slowly with detuning, in particular for detunings greater than $-8$~GHz, where $|\Delta\phi|<2\pi$.  For a linearly polarised pulse centred in this region with GHz bandwidth, the phase variation across its spectrum is sufficiently low so that the polarisation of the pulse as a whole can be rotated by a large angle.  This broadband rotation is demonstrated in figure~\ref{fig:pulserot}, where a 1~GHz bandwidth pulse polarised in the x direction is rotated into the y direction by increasing an applied magnetic field from zero to 230~G.  This $\pi/2$ rotation is accompanied by low absorption (less than 2\%) and no distortion since the pulse is many Doppler-broadened linewidths from resonance.  Light switching with such large bandwidth and with low loss/distortion has not been observed previously using an all-optical atom-light interface.    

\section*{Discussion}

For linearly polarised  continuous-wave light the plane of polarisation is rotated by an angle dependent on the medium's relative phase index, whilst its spectral sensitivity is dependent on the relative group index. These quantities can differ by more than six orders of magnitude, making the phase shift highly frequency dependent and hence of interest in the fields of interferometry and polarimetry.  The Faraday rotation angle was seen to go from zero at large detunings to $15\pi$ radians close to resonance, over a frequency range of $\sim20$~GHz.  This rotation corresponds to a Verdet constant of $3\times10^{-2}$~rad~G$^{-1}$~cm$^{-1}$, three orders of magnitude larger than in typical commercial Faraday rotators~\cite{Budker02}.  Large rotations are also seen in thin magneto-optical photonic crystals~\cite{Kahl04}, but the effect in such a system cannot be varied dynamically.  The rotation in the atomic vapour is temperature, magnetic-field and frequency dependent, therefore the effect can be used as a tunable polarisation switch.  In addition, we mention how polarimetry provides a useful spectroscopy tool~\cite{Pearman02}. The large bandwidth of off-resonant polarimetry could provide a useful technique to probe fast dynamics of Rydberg states in a thermal vapour~\cite{Mohapatra07,Mohapatra08} or atomic beam~\cite{Matt07} in the high density regime where interactions and hence novel non-linear optical effects are significant.  

For broadband pulses, with spectral widths of hundreds of MHz to GHz, in the wings of a resonance, the Faraday effect manifests itself as an asymmetry in the propagation of left/right circularly polarised light.  If the linearly polarised pulse's bandwidth is narrow enough such that the relative phase shift across it is essentially constant, the polarisation angle is rotated in the same way as for cw light.  Far from resonance, a GHz bandwidth pulse can be rotated by $\pi/2$~rad with less than 2\% loss and no distortion.  Increasing the phase variation across the pulse, e.g. by changing the applied field or carrier detuning, changes a single rotated linear pulse to split circular pulses in a dynamic way.  Splitting a pulse into two components separated by more than their temporal widths could be used for pump-probe experiments; or for photon switching or atomic ensemble entanglement~\cite{Polzik08} in quantum information applications. 

At higher temperatures and magnetic fields, the available bandwidth of many GHz make it possible for pulses of hundreds of picosecond duration to be delayed and rotated with high transmission.  Appropriate choice of atomic species and laser wavelength opens up the possibility of submarine (at $\sim450$~nm)~\cite{Menders91} and fibre-optic (at $\sim1.53$~$\mu$m) communications at the large bit rates required for rapid data transfer.  

\section*{Outlook}

The slow-light Faraday effect allows dynamic control of pulse shape and polarisation. Since this control is implemented via the applied magnetic field fast switching times of 1 microsecond or less would be difficult to achieve. However, the asymmetry in left and right circular polarisations due to the magneto-optical effect is not unique: any phenomenon which causes a change in the dispersive properties of a medium can lead to circular birefringence.  An alternative method is to redistribute population in the magnetic sublevels by optical driving.  A distribution which produces different coupling strengths for the two circular polarisations leads to the same effects described in this paper.  Augmenting (or replacing) the Faraday effect with optical control will potentially allow faster reconfiguration times than are possible via alteration of the applied magnetic field.  Hence any tunable device based on optical control would have faster switching times.  The large bandwidth of the slow light makes possible the probing of atomic dynamics on a timescale much shorter than the excited state lifetime.  The application of the effect to probe the coherent evolution of strongly driven thermal Rb atoms will be the focus of future work.

\section*{Method}
\subsection*{Theoretical Analysis}
\subsubsection*{Interferometry}

The Faraday effect arises due to the interference between the left and right circularly polarised components of the electric field.  In general, the phase shift, $\Delta\phi$, between two waves is produced by a differing optical path length, $\Delta(Ln)$, where $n$ is the phase index of the wave with velocity $v_p=c/n$, and $L$ is the length of the path taken by the wave.  The phase shift between two waves of angular frequency $\omega$ is given by~\cite{Boyd07cw}
\begin{equation}
\Delta\phi=\Delta (Lk) = \frac{\omega}{c}\Delta(Ln),
\label{eq:phase}
\end{equation}
where the wavenumber, $k$, and refractive index are taken to be real.  For the waves to be in-phase, one wave is delayed by an integer number of wavelengths with respect to the other.  The optical path difference is thus $\Delta(Ln)=m\lambda=m 2\pi c/\omega$, therefore $\Delta\phi=m2\pi$.  The separation in frequency between in-phase modes is known as the fringe spacing. 

For an interferometer based on Faraday rotation we analyse the relative phase of two orthogonal polarisations of light co-propagating through the same medium. The relative optical path length is thus $\Delta(Ln)=L\Delta n$, where the $\Delta n$ is the relative refractive index.

In dispersive media, $n$ is a function of $\omega$.  A change in frequency results in a change of phase due to a change in the quantity $\omega\Delta n$.  For the case $\Delta(Ln)=L\Delta n$, this change in phase is dependent on frequency via
\begin{eqnarray}
\delta\Delta\phi (\omega,L) & = & \Delta\phi (\omega+\delta\omega,L)-\Delta\phi (\omega,L) = L[\Delta k(\omega+\delta\omega)-\Delta k(\omega)]\nonumber\\
& = & L[\Delta k(\omega)+\frac{\rm{d}\Delta k(\omega)}{\rm{d}\omega}\delta\omega+\frac{1}{2!}\frac{\rm{d^2}\Delta k(\omega)}{\rm{d}\omega^2}\delta\omega^2+\ldots-\Delta k(\omega)]\nonumber\\
& = & \frac{L}{c}[\Delta n_g(\omega)+\frac{1}{2!}\frac{\rm{d}\Delta n_g(\omega)}{\rm{d}\omega}\delta\omega+\ldots]\delta\omega.
\label{eq:dphase}
\end{eqnarray}  
Here $\Delta n_g=n_g^+-n_g^-$ is the relative group refractive index, and $\rm{d}\Delta n_g/\rm{d}\omega$ is the relative group velocity dispersion (GVD).  Higher order terms in the Taylor expansion lead to distortion of optical pulses. For the case of low GVD the expansion in (\ref{eq:dphase}) can be taken to first order, hence $\delta\Delta\phi=(L\Delta n_g /c)\delta\omega$ with a corresponding fringe spacing of $\delta\omega=2\pi c/L\Delta n_g$.  In regimes where GVD is not negligible, the group index changes with frequency and hence the fringe spacing is not constant at frequencies where the $n_g$ is dominated by higher order terms.

In the limit that $\delta\omega \rightarrow 0$, equation (\ref{eq:dphase}) becomes
\begin{eqnarray}
\frac{\rm{d}(\Delta\phi)}{\rm{d} \omega} & = &  \frac{L}{c}\Delta n_g.
\label{eq:dp}
\end{eqnarray} 
A small change in frequency thus produces a change in relative phase difference proportional to the relative group refractive index.  In a slow-light medium the spectral sensitivity described by equation~(\ref{eq:dp})  scales as $n_g$, whereas the sensitivity scales as $n$ for a non-dispersive medium.  Since the group index can be several orders of magnitude larger than the phase index, the sensitivity of the interferometer is correspondingly enhanced.  Note that this large sensitivity relies upon a change in the quantity $\omega\Delta n(\omega)$.       

\subsubsection*{Balanced polarimetry}

In order to determine the relative phase shift between the circular components of light we use a balanced polarimeter~\cite{Huard}.  We set a polarisation beam splitter at $\pi/4$~rad to the linearly polarised light such that in the absence of any optical rotation there is an equal amount of vertical and horizontal light incident on separate photodetectors.  By subtracting the two signals we are left with a zero background such that a small change in the amount of light in either linear polarisation can be detected.  For an input polarisation $\textbf{\rm{e}}=\frac{1}{\sqrt{2}}(\hat{\textbf{\textrm{x}}}+\hat{\textbf{\textrm{y}}})$, the output intensity signals for an initial intensity $I_0=I_{x_0}+I_{y_0}$ are~\cite{Pearman02}
\begin{eqnarray}
I_x-I_y &=& I_0\sin(\Delta\phi) \textrm{e}^{-(\alpha^++\alpha^-)L/2},\label{eq:polar1}\\
I_x+I_y &=& \frac{1}{2}I_0(\textrm{e}^{-\alpha^+L}+\textrm{e}^{-\alpha^-L}).
\label{eq:polar2}
\end{eqnarray}
Notice in equation~(\ref{eq:polar1}) that the period of oscillation does not depend upon any attenuation due to the absorption coefficients $\alpha^\pm$: the differential signal is sensitive to optical rotation whilst being insensitive to ellipticity induced via circular dichroism. 

\subsection*{Experimental Method}

\begin{figure}[tbh]
\centering
\includegraphics[width=8.5cm]{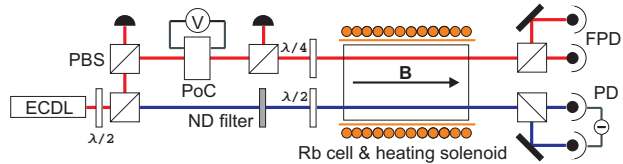}
\caption{Schematic diagram of the experimental apparatus.  The output of an external cavity diode laser (ECDL) is split with a polarisation beam splitter (PBS) to provided both cw (blue) and pulsed (red) light.  Optical pulses are generated using a Pockels cell (PoC) supplied with a pulsed voltage.  The cw light is attenuated with a neutral density (ND) filter.  Waveplates ($\lambda/2$ and $\lambda/4$) are used to control the polarisation of both beams before they pass through the vapour cell.  The two orthogonal linear components of the pulse are collected on separate fast photodiodes (FPD), whilst the two components of the cw beam are collected on a differencing photodiode (PD).}
\label{fig:setup}
\end{figure}

The experimental setup is seen in figure~\ref{fig:setup}.  The source of both the cw and pulsed light was a Toptica DL100 external cavity diode laser (ECDL) with wavelength 795.0~nm.  The beam has a 1/e$^2$ radius of $(2.00\pm0.05)$~mm.  The light was then split using a $\lambda/2$ plate and polarisation beam splitter cube (PBS). Most of the light was used to produce an optical pulse, a small fraction was used for the cw spectroscopy. The $(1.5\pm0.2)$~ns FWHM pulse has a peak intensity of 4~mW/mm$^2$.  The pulse profile is approximately Gaussian in time, and therefore has a spectral FWHM of $\sim300$~GHz and bandwidth of $\sim1$~GHz; the bandwidth is here defined as the frequency range containing 99\% of the pulse spectrum.  The pulse was produced using a Pockels cell (PoC) placed between two crossed polarisers, which was driven by a pulse generator with a repetition rate of 50~Hz.  The polarisations of the separate beams were then adjusted with waveplates before entering the vapour cell, and the cw beam was attenuted to 30~nW/mm$^2$ such that it was in the weak-probe limit~\cite{Siddons08}.  The 75~mm fused silica cell contained a mixture of the two naturally occuring isotopes of Rb according to the ratio $^{87}$Rb:$^{85}$Rb 99:1.  The cell windows were angled to prevent etalon fringes due to back-reflection. The cell was placed in a solenoid heating unit, based on the design from reference~\cite{Danny}.  The current produced both the magnetic field and Joule heating required for the experiment; data requiring zero applied field were taken by briefly turning off the current.  The variation in magnetic field along the cell is about 10\%.  No attempt was made to shield the apparatus from the laboratory field. 

\section*{Acknowledgements}

The authors would like to thank M. P. A. Jones for valuable discussion.  This work was funded by EPSRC. 

\section*{Author contributions}

P.S. undertook the experiment and theoretical modeling, and contributed in the writing of the paper. N.C.B. and Y.C. assisted with the experiment. C.S.A. and I.G.H contributed to the writing of 
the paper, and were responsible for the project management.

\end{document}